# Characterizing the current systems in the Martian ionosphere

Jiawei Gao[1,2], Shibang Li[3], Anna Mittelholz[4], Zhaojin Rong[1,2], Moa Persson[5], Zhen Shi[1,2], Haoyu Lu[3], Chi Zhang[6], Xiaodong Wang[7], Chuanfei Dong[6], Lucy Klinger[8,9], Jun Cui[10], Yong Wei[1,2], Yongxin Pan[1,2]

[1]Key Laboratory of Earth and Planetary Physics, Institute of Geology and Geophysics, Chinese Academy of Sciences, Beijing, China

[2]College of Earth and Planetary Sciences, University of Chinese Academy of Sciences, Beijing, China

[3]School of Space and Environment, Beihang University, Beijing, China

[4]Department of Earth Sciences, ETH Zurich, Zurich, Switzerland

[5]Swedish Institute of Space Physics, Uppsala, Sweden

[6]Center for Space Physics and Department of Astronomy, Boston University, Boston, MA, USA

[7]Swedish Institute of Space Physics, Kiruna, Sweden

[8]Fudan University, Shanghai, China

[9]The Shanghai Institute for Mathematics and Interdisciplinary Sciences, Shanghai, China

[10]Planetary Environmental and Astrobiological Research Laboratory, School of Atmospheric Sciences, Sun Yat-sen University, Guangdong, China

Corresponding author: gaojw@mail.iggcas.ac.cn; rongzhaojin@mail.iggcas.ac.cn

**Abstract**

When the solar wind interacts with the ionosphere of an unmagnetized planet, it induces currents that form an induced magnetosphere. These currents and their associated magnetic fields play a pivotal role in controlling the movement of charged particles, which is essential for understanding the escape of planetary ions. Unlike the well-documented magnetospheric current systems, the ionospheric current systems on unmagnetized planets remain less understood, which constrains the quantification of electrodynamic energy transfer from stars to these planets. Here, utilizing eight years of data from the Mars Atmosphere and Volatile EvolutioN (MAVEN) mission, we investigate the global distribution of ionospheric currents on Mars. We have identified two distinct current systems in the ionosphere: one aligns with the solar wind electric field yet exhibits hemispheric asymmetry perpendicular to the electric field direction; the other corresponds to the flow pattern of annually-averaged neutral winds. We propose that these two current systems are driven by the solar wind and atmospheric neutral winds, respectively. Our findings reveal that Martian ionospheric dynamics are influenced by the neutral winds from below and the solar wind from above, highlighting the complex and intriguing nature of current systems on unmagnetized planets.

**1 Introduction**

Unlike Earth, Mars lacks a global dipolar magnetic field. The solar wind plasma and the frozen-in interplanetary magnetic field (IMF) interact directly with the Martian highly conductive ionosphere, inducing associated currents and forming an induced magnetosphere[1,2,3]. The IMF, draping around the ionospheric obstacle, is stretched by the solar wind and forms an induced magnetotail, as is the case for Mars, Venus, Titan, Pluto, and many comets and exoplanets[4,5,6]. The currents and their associated magnetic fields play a crucial role in deflecting plasma around Mars and shielding the planetary ionosphere from solar wind. The transfer of energy and momentum from the Sun to the Martian atmosphere results in a significant part of atmospheric plasma, or ionized gas, to escape the planet's gravity[7,8,9]. Consequently, understanding the electrodynamics of the induced magnetosphere is crucial for unraveling the history of planetary climate and the evolutionary path of habitability.

The global current systems in the induced magnetosphere of Mars have been studied, revealing their dominant structure[10,11]. When the solar wind encounters a planetary magnetic obstacle, the discrepancy in gyrations between solar wind ions and electrons generates Chapman-Ferraro-type currents at the bow-shock and the induced magnetosphere boundary (IMB)[11,12]. The bow shock and IMB currents flow in the opposite direction to the solar wind electric field and, in turn, drive induced currents in the Martian ionosphere. The magnetotail currents, which can be considered as tailward extension of the magnetopause currents, flow through the magnetotail current sheet along the direction of the solar wind electric field and connect with the IMB and bow-shock currents at the flanks[11,13]. Both the magnetopause and magnetotail current systems have been validated by recent computer simulation[14]. In addition to those two current systems, a sunward current in the magnetosheath and a tailward current in the entire magnetotail have been unveiled[11] (see Fig. 1a). A Previous study suggested that the sunward current in the Martian magnetosheath might be connected to the electric polar regions of the Martian ionosphere[11]. However, the exact nature of the linkage between these magnetospheric currents and the Martian ionosphere is still not well-defined, underscoring the need for a more detailed characterization of the ionospheric currents.

In contrast to Earth's well-known ionospheric current systems, including the solar quiet (Sq) current and the field-aligned currents (FACs), the distribution of currents in the Martian ionosphere is not well understood[15,16]. Previous studies based on data from the Mars Global Surveyor (MGS) spacecraft, mainly collected at 400 km altitude, that is above the ionospheric peak and at fixed local times, proposed that Martian ionospheric currents could generate a horizontal magnetic field, but the lack of global spacecraft coverage has limited data-based studies[17,18,19]. Similar to the Earth's Sq currents, the ionospheric currents on Mars were suggested to be driven by atmospheric neutral winds; however, a global map detailing these currents does not yet exist[20,21,22,23]. Fortunately, the MAVEN spacecraft has been providing extensive magnetic field and plasma measurements across ionospheric altitudes (120-500 km) at variable local times[24], thus enabling in-situ observations of the ionospheric electromagnetic environment (refer to Supplementary Fig. 1 for data distribution). Using MAVEN data, an electric northward current has been identified in the dayside ionosphere, flowing along the direction of the solar wind electric field[11]. The ionospheric current was also observed by the Mars InSight lander[25] and Zhurong rover[26], which recorded temporal variations of the Martian surface magnetic field[25,26,27]. Previous studies have shown that both atmospheric neutral winds[22,23] and the solar wind[11,28] are significant contributors to ionospheric dynamics and, consequently, the generation of ionospheric currents. Therefore, a comprehensive and detailed global map of Martian ionospheric currents at various altitudes is essential to understand and quantify their global distribution. Furthermore, characterizing Martian ionospheric currents contributes to our understanding of current distributions on other unmagnetized planets with ionospheres, such as Venus and Titan, suggesting a potentially universal phenomenon.

One of the main challenges in investigating ionospheric currents on Mars is the presence of Martian localized crustal magnetic fields[29,30]. Significant currents can be locally generated in regions with significant crustal field[31,32,33,34], which can substantially alter the magnetic topology at ionospheric altitudes. Since Martian crustal fields vary spatially, there are abundant regions where the ionospheric magnetic fields remain unaffected by the crustal field (refer to Supplementary Fig. 2). Therefore, to minimize the influence of crustal field, our study focuses only on datasets from these non-crustal field regions, specifically areas where the intensity of the crustal fields at 120 km altitude is less than 10 nT (see Method 4.1).

Here, based on a statistical analysis of eight years of MAVEN measurements from November 2014 to May 2022, we present a detailed map of the magnetic fields and currents at ionospheric altitude. We find that currents driven by both the solar wind and atmospheric neutral winds coexist within the Martian ionospheric dynamo regions. Our results could contribute to quantifying the energy transfer from the stellar to the planetary atmosphere, a process fundamental to understanding planetary atmospheric evolution.

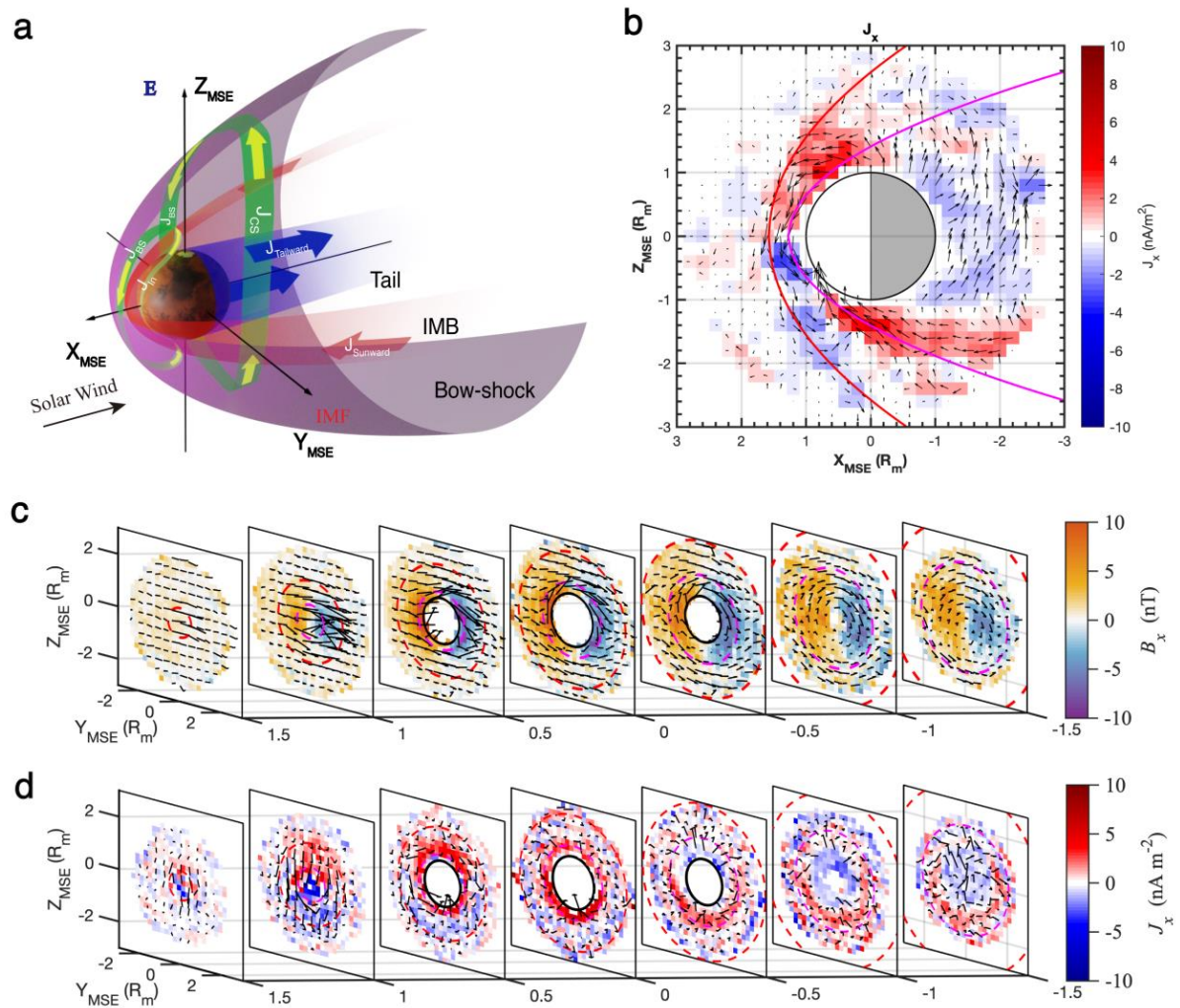

**Fig. 1. Illustration of the current systems in the induced magnetosphere of Mars.** (a) Diagram of the current systems in the induced magnetosphere. The magnetopause and magnetotail current system are colored green, with the yellow arrows indicating the current direction. This current pattern consists of the bow-shock current ($J_{BS}$), the magnetotail current sheet current ($J_{CS}$), the sunward/tailward currents ($J_{Sunward}$/$J_{Tailward}$), and the ionospheric induced currents ($J_{In}$). For clarity, the currents at the induced magnetospheric boundary that connect with the $J_{In}$ and $J_{CS}$ are not shown. (b) The current distribution in the $XZ_{MSE}$ plane, through the meridian plane with a thickness of 0.2 $R_m$, centered at $Y_{MSE} = 0$. (c-d) Magnetic field (c) and current (d) slice maps in $YZ_{MSE}$ plane along $X_{MSE}$ direction. In each slice, arrows represent the direction of magnetic field (c) and current (d) in the $YZ_{MSE}$ plane. The red and magenta dashed lines denote the shape of the bow shock and the induced magnetosphere boundary[97]. The black circle denotes the body of Mars.

## 2 Results

### 2.1 Magnetic field and current distribution in the Martian magnetosphere

To present a comprehensive view of the global current systems, we begin by re-examining the current distributions in the Martian magnetosphere using a more extensive dataset than those used in previous studies[11]. Since the magnetic field geometry of the Martian induced magnetosphere is controlled by the orientation of the upstream IMF[35,36,37], we first rotated the magnetic field data from the Mars-Solar-Orbital (MSO) coordinates to the Mars-Solar-Electric (MSE) coordinates (see Method 4.1). In the MSO coordinates, the X-axis points from Mars to the Sun, the Z-axis points toward orbital north, and the Y-axis is opposite to Mars' orbital motion. In MSE coordinates, X-axis is anti-parallel to the solar wind flow, the Z-axis aligns with the solar wind electric field ($\boldsymbol{E}_{SW} = -\boldsymbol{v}_{SW} \times \boldsymbol{B}_{IMF}$, where $\boldsymbol{v}_{SW}$ is the solar wind flow and $\boldsymbol{B}_{IMF}$ is the IMF in solar wind), and the Y-axis completes the right-handed coordinate system (see Supplementary Fig. 3 for the coordinates). Lastly, we calculated the current distribution by taking the curl of the magnetic field.

Fig. 1c displays the slices of magnetic field distribution in the $YZ_{MSE}$ plane at varying distances from upstream to downstream, while Fig. 1d shows the corresponding slices of calculated current density. The IMF begins to drape around the planet as the solar wind encounters the ionosphere obstacle, and the opposite polarities of the magnetic field $B_x$ in $\pm Y_{MSE}$ hemisphere consistently demonstrate the draping field lines of the induced magnetosphere[37,38]. Additionally, a clockwise rotating magnetic field, viewed from the Sun towards Mars and extending from the terminator to the magnetotail, is clearly observed[39,40,41,42]. This rotating magnetic field was previously attributed to the sunward and tailward currents in the magnetotail[11]. The magnetopause current system on the dayside and the magnetotail current system are distinctly visible in Fig. 1b.

Several potential connections exist between the magnetospheric and ionospheric currents. The magnetopause currents could connect with ionospheric currents on both the dayside and the flanks, as indicated by a reversal in $J_Z$ signal (Supplementary Figs. 4). Currents at the bow shock and IMB flowing toward the $-Z_{MSE}$ direction (electric southward) are likely closed with ionospheric currents in the $+Z_{MSE}$ direction (electric northward). Meanwhile, the tailward currents in the magnetotail are likely connected to the nightside ionosphere, although a direct connection is not clearly visible.

## 2.2 Magnetic field distribution in the Martian ionosphere

The characterization of ionospheric currents on Mars depends on understanding the distribution of the Martian ionospheric magnetic field, which is driven by two main factors: the induction and transportation of the IMF carried by the solar wind, and heating by solar radiation. Although MSE coordinates effectively capture the main features of the Martian induced magnetosphere, it may smear out any features that vary with solar local time by rotating the spacecraft's position along the X-axis[43]. In other words, the MSE frame is suitable for investigating phenomena governed by the upstream solar wind IMF orientation, while the MSO frame is better suited for analyzing phenomena driven by solar irradiation heating. Consequently, it is necessary to employ both coordinate systems, i.e., MSE and MSO, to accurately display the magnetic field distributions. After statistically analyzing the ionospheric magnetic field distribution in both MSO and MSE coordinates, we calculated the current density by taking the curl of the magnetic field within a spherical shell spanning ionospheric altitudes from 150 to 500 km, partitioned by bins with width of 50 km (see Method 4.2).

Fig. 2a-f and Fig. 3a-f show maps of the ionospheric magnetic field in the MSE and MSO coordinates, respectively (see Supplementary Figs. 5-6 for additional slices at various ionospheric altitudes). In the MSE coordinates, the IMF draping around Mars is observed down to an altitude of 150 km, consistent with MHD simulation results[44] and substantially lower than previously observation estimated[45,46,47] (Fig. 2f). The ionospheric magnetic field primarily exhibits a horizontal distribution. However, we observed that the magnetic fields possess an electric northward ($-B_\theta$) and a southward ($+B_\theta$) component in the $-Y_{MSE}$ and $+Y_{MSE}$ hemisphere, respectively, with magnitudes up to 20 nT. This distribution pattern suggests a clockwise-rotational field structure when viewed from above the ionosphere.

In the MSO coordinates, the IMF's draping pattern is not evident. The occurrences of both positive and negative $B_y$ components of the upstream IMF are nearly equal[48] (Supplementary Fig. 7), which results in the draping features being averaged out. Instead, a magnetic field structure with clockwise-rotation is present in both the dayside ionosphere and terminator regions. Furthermore, an inward magnetic field ($-B_r$) in the Northern hemisphere and an outward magnetic field ($+B_r$) in the Southern hemisphere is visible (Fig. 3d). The presence of a clockwise rotating magnetic field, coupled with the inward and outward magnetic field on the dayside, suggests the existence of ionospheric currents driven independently of the solar wind.

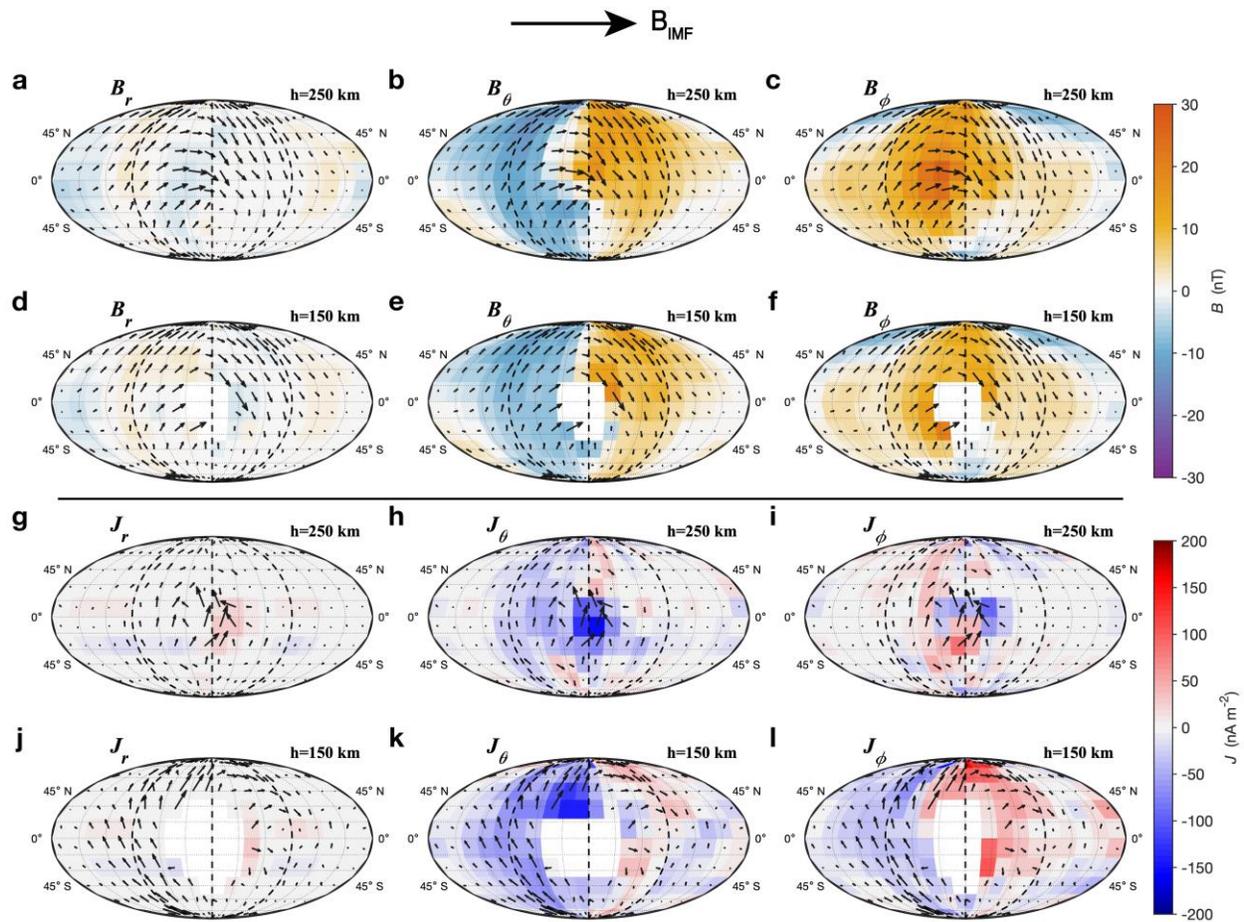

**Fig. 2**. **The distributions of magnetic field and current in the Martian ionosphere in the MSE coordinates.** Panels from top to bottom display the distributions of magnetic field (a-f) and current (g-i) at altitudes of 250 km and 150 km in the Mollweide projection, respectively. Panels from left to right show the radial $B_r$ ($J_r$), southward $B_\theta$ ($J_\theta$), and eastward $B_\varphi$ ($J_\varphi$) component of the magnetic field (current), respectively. Arrows in each panel indicate the directions of the horizontal

magnetic field (current) components. The vertical dashed lines in each panel denote the terminator. The gap in the panels at 150 km altitude indicate the absence of measurements in those area.

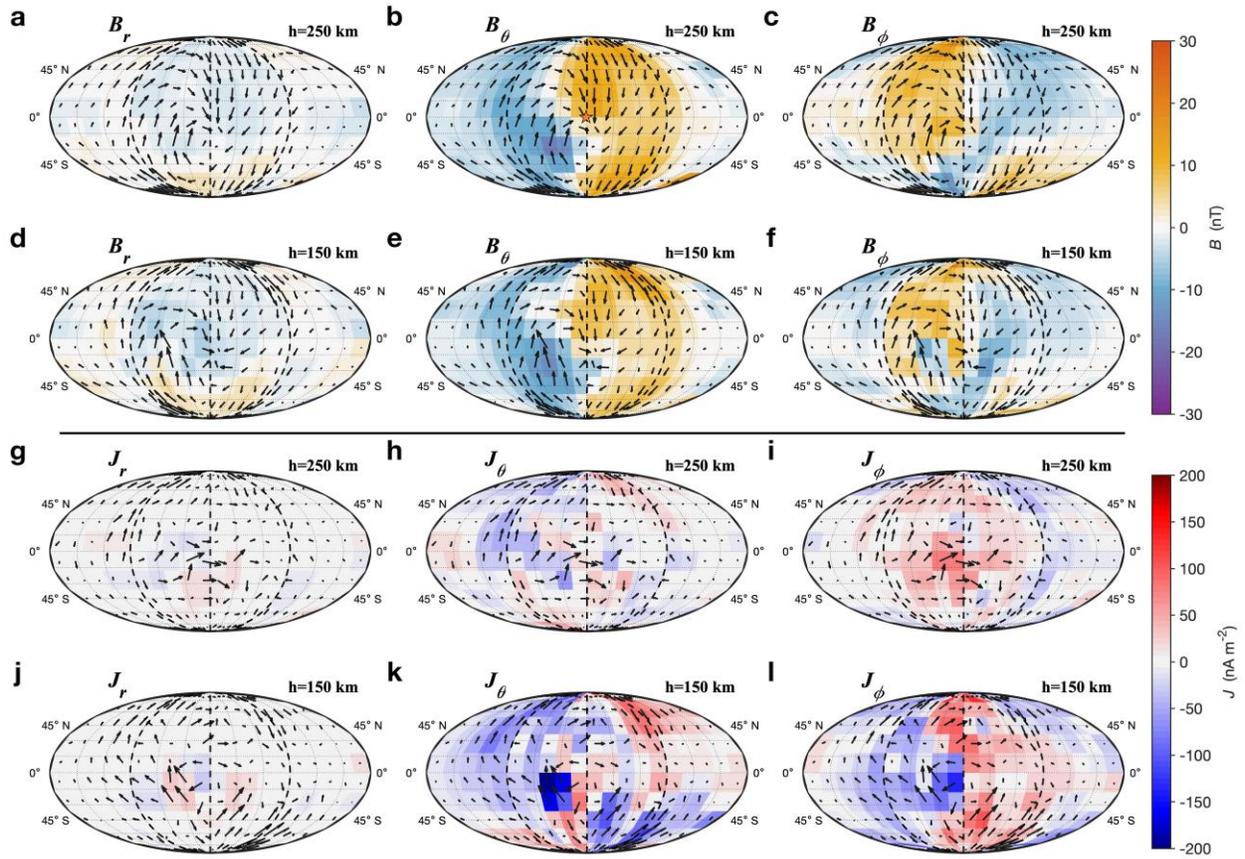

**Fig. 3. The distribution of magnetic field and current in the Martian ionosphere in the MSO coordinates.** The format is the same as that of Fig. 2.

### 2.3 Current distribution in the Martian ionosphere

The ionospheric current distribution in the MSE and MSO coordinates are depicted in Fig. 2g-l and Fig. 3g-l, respectively (see Supplementary Figs. 8-9 for slices at other altitudes). It is evident that in both MSE and MSO coordinates, the minor radial component of the current ($|J_r| < 20$ nA/$m^2$) indicates that the ionospheric currents at lower altitudes predominantly flow horizontally (Fig. 2j and Fig. 3j). For convenience, we label the current patterns at an altitude of 150 km in the MSE coordinate as $J_{In}$ and in the MSO coordinate as $J_{Sq}$.

In the MSE coordinates, within the 150 km to 250 km altitude range, $J_\theta$ and $J_\varphi$ show an enhancement as altitude decreases. At an altitude of 150 km, a noticeable electric northward current ($-J_\theta$) is present on the dayside, reaching a maximum of 150 nA/$m^2$. An electric northward (southward) current is observed around the $-Y_{MSE}$ ($+Y_{MSE}$) terminator regions, each with a magnitude of approximately 50 nA/$m^2$ (Fig. 2k). Concurrently, a $-J_\varphi$ ($+J_\varphi$) is evident in the $-Y_{MSE}$ ($+Y_{MSE}$) hemisphere, contributing to a tailward current (Fig. 2l).

In the MSO coordinates, the current distribution exhibits an irregular pattern at higher altitudes, spanning from 300 km to 500 km. At 250 km altitude, an eastward current is observed around the equatorial region. At 150 km altitude, the dayside current system consists of two flows: one in each Northern/Southern hemisphere, each extending from dawn to dusk regions. These two current flows are characterized by a quadrupole pattern of $J_\theta$ (Fig. 3k) and a bipolar pattern of $J_\varphi$ (Fig. 3l) around the noon meridian.

## 3 Discussion

### 3.1 Formation mechanism of the ionospheric current systems

In the Martian ionosphere, two principal current systems, $J_{In}$ and $J_{Sq}$, are identifiable in the MSE and MSO coordinates, respectively. $J_{In}$ represents the electric northward ionospheric induced current, linking the electric southward bow-shock and IMB currents[11]. Previous research has suggested that $J_{In}$ acts as a load current driven by an electric potential difference generated by the charges flow from bow-shock currents[10,11]. Our findings reveal that $J_{In}$ extends down to an altitude of 150 km. However, we observed an unexpected asymmetry in $J_{In}$ across the $Y_{MSE}$ hemisphere. Specifically, $J_{In}$ is electric-northward in the $-Y_{MSE}$ hemisphere but exhibits an irregular pattern in the $+Y_{MSE}$ hemisphere. The underlying mechanism of this hemispherical asymmetry is not yet understood. We tentatively hypothesize the existence of an additional current system, flowing clockwise at the terminator in the $YZ_{MSE}$ plane, which superimposes on $J_{In}$ and causes this asymmetry. Such a clockwise rotating current could generate a sunward magnetic field above its source region, resembling induced currents with diamagnetic properties.

To further investigate the hemispherical asymmetry of the $J_{In}$ current, we compared MAVEN observations with results from 3D Multi-fluid Hall-MHD simulation (details provided in Method 4.3). Notably, our simulation results (Fig. 4) did not replicate the asymmetry observed in the $\pm Y_{MSE}$ hemispheres (Supplementary Figs. 10). Additionally, a recent hybrid simulation also failed to replicate the observed current asymmetry[14]. Since these simulations didn't consider any ionospheric electrodynamic processes[50,51,52], the failure to replicate the asymmetry by those simulations may suggest that the asymmetry of $J_{In}$ is more likely an intrinsic feature of ionospheric currents rather than being driven by the magnetospheric processes.

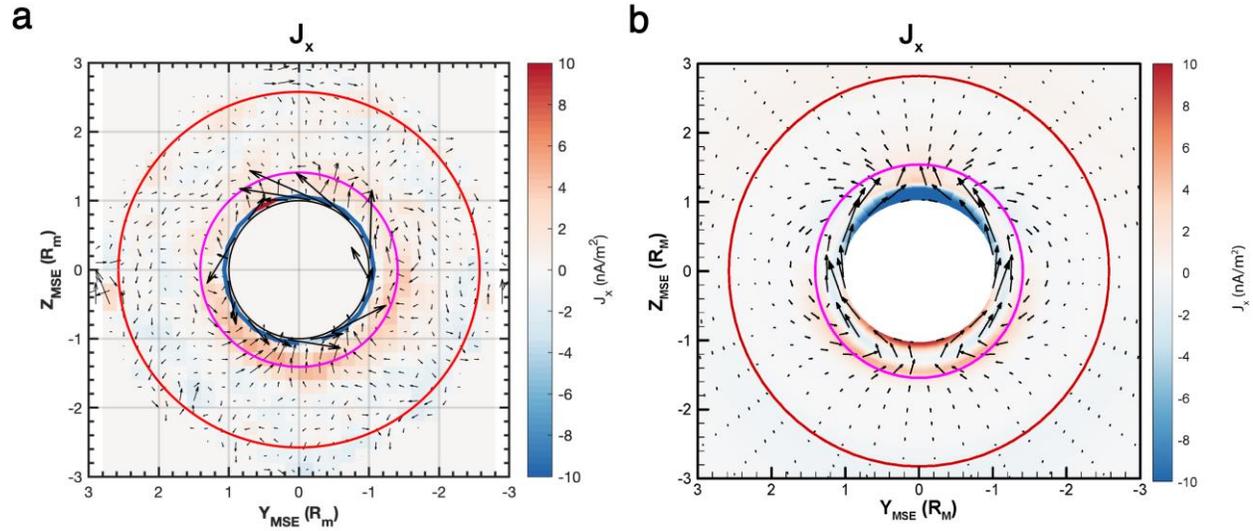

**Fig. 4**. **Slices of the current distribution in the induced magnetosphere of Mars for both MAVEN observation and simulation.** The current pattern derived from (a) MAVEN observations and (b) a Multi-fluid Hall-MHD simulation in the terminator plane, as seen from the magnetotail. Color represents the strength of $J_x$, while the arrows represent the magnitude and direction of the currents in the projected plane. In Panel a, the ionospheric current at 150 km altitude is superposed on the magnetospheric current distribution map. The red and magenta circles denote the shape of the bow shock and the IMB, respectively[97].

$J_{Sq}$ denotes the Solar quiet (Sq) current of the Mars ionosphere. At an altitude of 150 km on the dayside, the quadrupole pattern of $J_\theta$ and the bipolar pattern of $J_\varphi$ about the noon meridian imply that $J_{Sq}$ consists of two current patterns, each located in the Northern or Southern hemispheres (Fig. 3k-l). In the high latitudes of both the Northern and Southern hemispheres, $J_{Sq}$ flows from dawn to dusk (Fig. 5). Near the dawn region, $J_{Sq}$ exhibits northward components in the Northern hemisphere, while converging in the equatorial region at dusk. Analogous to Earth's Sq current system, the Martian $J_{Sq}$ current system could be similarly powered by the atmospheric neutral winds[16,22,23].

In the ionospheric dynamo region, at approximately 150 km altitude (Supplementary Fig. 11), ions are coupled to the neutral wind through collisions, i.e., their collision frequency with neutrals is higher than their gyration frequency, while the opposite is true for electrons. The ions are coupled with neutral winds and electrons gyrate about the penetrated magnetic field lines of IMF, leading to the generation of ionospheric currents through the ion movement. Consequently, considering approximately equal densities of ions and electrons, the current density $\boldsymbol{J} = n_e q \boldsymbol{v}_n$ is proportional to the velocity of the neutral wind, where $n_e$ is the electron density, $q$ is the elementary charge, and $\boldsymbol{v}_n$ is the wind velocity. Assuming the peak electron density in the dayside ionosphere is approximately $10^{11}/m^3$ [53] and significantly decreases on the nightside[54], we can deduce the annual average of the neutral wind patterns in the MSO coordinate from the current distribution by $\boldsymbol{v}_n = \boldsymbol{J}/n_e q$. As shown in Fig. 5, the derived wind patterns align well, both in magnitude and direction, with the annual averaged neutral winds predicted by a Mars Global Circulation Model (MGCM) (see Method 4.4). This evidence supports that these currents are driven by the neutral winds.

Note that discrepancies still exist between the wind fields estimated from the current density and those predicted by the MGCM. First, we notice data points with pronounced northward directions near the equator between local times 8 h and 10 h. The calculated values of these data points are significantly stronger than that those predicted by the MGCM model. However, these data points exhibit considerable uncertainty, indicating that the observed strong currents in this region may contain substantial noise and should not be over-interpreted (see Method 4.5 for uncertainty estimation). Second, in the polar regions at local times 18 h, the wind velocity estimated from the current density exceed those predicted by the MGCM model. This deviation could be attributed to either the uneven data distribution across seasons or the residue of the draped IMF in the MSO coordinate not being fully averaged out. It is worth noting that the instantaneous wind velocity in the MGCM model[55] at a given time could be much higher than the average wind velocity we estimated from the current density.

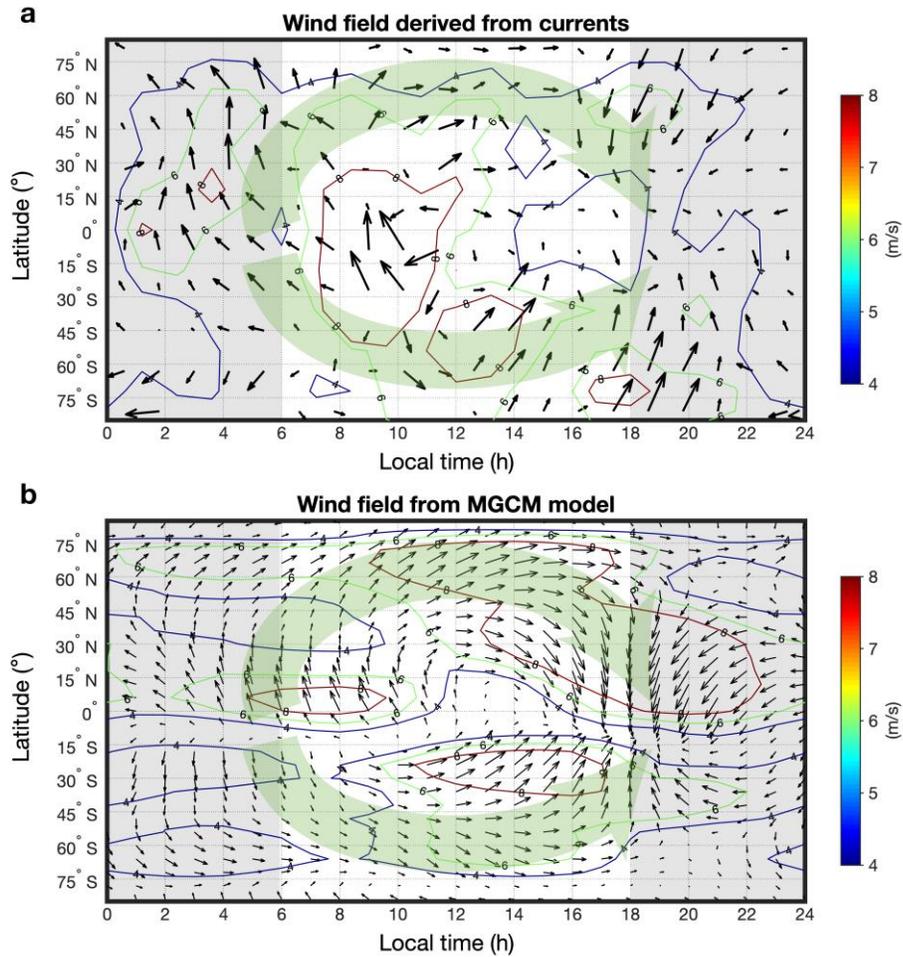

**Fig. 5**. **Neutral wind field pattern inferred from ionospheric current and the MGCM model**. (a) The horizontal wind field estimated from the ionospheric current at the altitude of 150 km. The neutral wind velocity $v_n$ is calculated using the formula $v_n = J/n_e q$, where $J$ is the ionospheric current, $n_e$ is the electron density, and $q$ is the elementary charge. The X-axis represents the East-West direction, and the Y-axis represents the North-South direction. Arrows indicate the direction and the magnitude of the wind field in the horizontal plane. The shaded area denotes the nightside regions. (b) The horizontal wind field at 130 km altitude predicted by Mars Global Circulation Model. The wind field is averaged across all Martian seasons and geographic longitudes[22].

The driving mechanism of the ionospheric current can be further corroborated by examining the distribution of the ionospheric electric field (Fig. 6), which is calculated based on the current and conductivity distribution (Supplementary Fig. 12) and governed by Ohm's law. In the MSE coordinates, the electric field predominantly points northward on the dayside, with magnitudes reaching up to $10^6$ nV/$m$, aligning with previous estimates of the ionospheric electric field magnitude[49]. These observations support the hypothesis that the current in the MSE coordinates is mainly induced by the electric potential difference between the electric South pole and North pole. In the MSO coordinates, the dayside ionospheric electric field exhibits two loop-like flows, aligning with the patterns of ionospheric currents. In the ionospheric dynamo region, Ohm's law defines the current density **J** as $\mathbf{J} = \sigma(\mathbf{E} + \mathbf{U} \times \mathbf{B})$. Given that the term $\mathbf{U} \times \mathbf{B}$ is negligible compared to **E** (Supplementary Fig. 13), the ionospheric currents are driven by electric field, which are most likely powered by neutral winds.

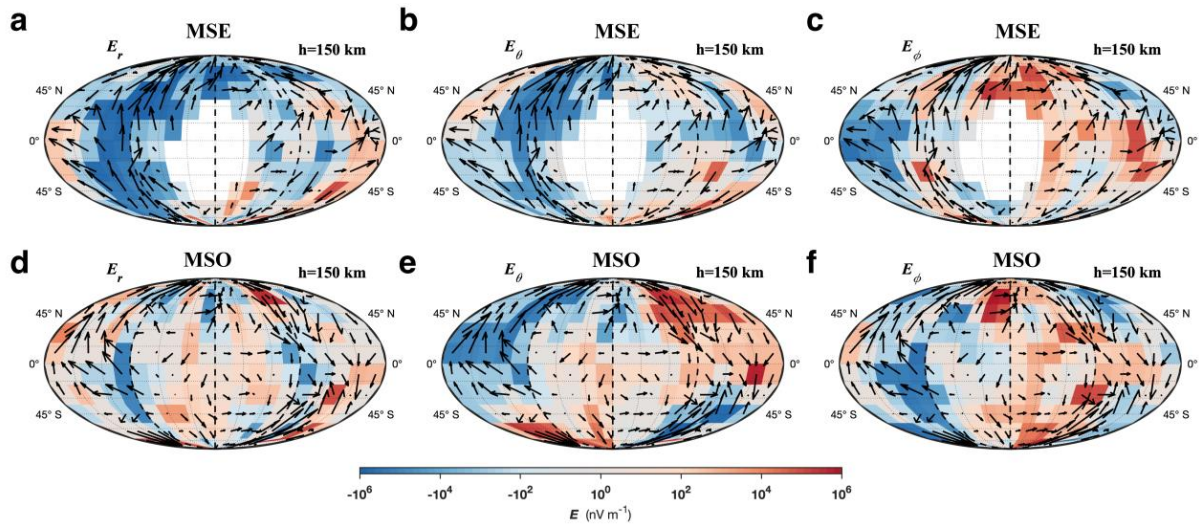

**Fig. 6**. **Electric field distribution in the Martian ionosphere.** The $E_r$, $E_\theta$, and $E_\varphi$ component of the ionospheric electric field at 150 km altitude in the MSE (a-c) and MSO (d-f) coordinates. Arrows represent the amplitude and direction of the electric field in the horizontal plane, with length represented logarithmically.

### 3.2 Comparison of the current systems on Earth and Mars

The Sq current system is a well-established feature of the Earth's ionosphere, and similar phenomena are also observed on Mars within distinct magnetic environments. Intriguingly, the driving mechanisms for the Sq current systems on both planets appear to be similar: the motion of the neutral atmosphere generates the ionospheric dynamo[23,56,57].

Both Earth's and Mars' Sq currents are situated within an altitude range of 100-150 km, known as the dynamo region, where electron density peaks[53,58]. The Pedersen and Hall conductivities in Earth's ionosphere are one to two orders of magnitude weaker than those of Mars[20,59]. Notably, the current intensity of Earth's Sq current system can reach several $\mu A/m^2$, and is one to two orders of magnitude greater than that of Mars' Sq current, with a maximum current density of about 150 $nA/m^2$. As a result, the typical electric field magnitudes in Earth's dynamo region often reach several mV/m, which is about three orders of magnitude greater than those on Mars, where values are typically in the range of several µV/m on the dayside. We also estimated the rate of Joule heating ($\boldsymbol{J} \cdot \boldsymbol{E}$), a phenomenon occurring when current passes through a conductive material, in the Martian ionosphere (Supplementary Fig. 14). In the MSO coordinates, Joule heating rates can reach up to $10^{-11}$ Joule $\cdot s^{-1}m^{-3}$ in the terminator and midnight regions, while in the subsolar regions, they are approximately $10^{-14}$ Joule $\cdot s^{-1}m^{-3}$. The comparable Joule heating rates in both MSE and MSO coordinates suggest that both solar wind and neutral winds are significant contributors to ionospheric dynamics.

The Earth's Sq current system is modulated by seasons[58]. Similarly, seasonal modulation is observed in the Martian Sq currents. While the pattern of the Sq current system remains roughly unchanged throughout the Mars year, the current density is generally stronger in the Northern hemisphere during Northern hemisphere summer (Supplementary Fig. 15), aligning with the seasonal variations seen in Earth's Sq currents. Increased solar radiation results in higher plasma densities in the summer hemisphere[60], potentially amplifying ionospheric currents. Furthermore, the occurrence of dust storms on Mars exhibits clear seasonal variation. Dust storm can elevate the peak-height of plasma density of ionosphere, potentially modulating the ionospheric currents[27,61]. In the future, improved data statistics could enable a more precise delineation of the effects of dust storms on ionospheric currents.

In Earth's high-latitude regions, field-aligned currents, also known as Birkeland currents, facilitate a connection between the magnetosphere and the polar ionosphere[62,63]. On Mars, we do not find significant vertical currents in the Martian ionosphere (Fig. 2j and Fig. 3j). In the MSE coordinates, we observed very weak inward and outward currents in the low-latitude regions on the dayside, potentially closing with the magnetopause current systems. Additionally, the tailward currents observed in the Martian ionospheric terminator region are likely connected to the tailward currents in the magnetotail. However, the tailward current in the ionosphere is confined to the regions around the terminator, despite its current density being significantly stronger than that in the magnetospheric tail. The curl-B technique used to calculate current has its own limitations, as it cannot resolve spatial current variations that are smaller than the bin size (50 km in the vertical, 18° in latitude and longitude). Thus, we cannot rule out the presence of significant currents in very thin layers of the Martian ionospheric boundary, for example, current densities at the IMB have been suggested to reach up to 200 nA/$m^2$ [28].

On Earth, neutral winds in the ionospheric dynamo region are predominantly driven by upward-propagating atmospheric tides originating from the lower atmosphere[64]. Specifically, the semi-diurnal tide emerges as the principal contributor to the wind field at this altitude range. A similar wind pattern is observed on Mars[65], suggesting that the semi-diurnal tides from the lower atmosphere may also serve as the primary driver for Martian ionospheric dynamo. The upward propagation of semi-diurnal tides may drive the eastward current observed at 250 km altitude, warranting further confirmation.

### 3.3 Influence of the ionospheric current systems on the Martian space environment

Currents play an important role in shaping the space environment by generating self-consistent magnetic fields and guiding the motions of charged particles. Here we qualitatively estimate the influence of the ionospheric current system on the Martian space environment.

$J_{In}$ is connected with the currents at bow-shock and IMB, forming a dayside magnetopause current system. This creates a strong magnetic pressure gradient, forming a magnetic barrier between the two current layers. The opposing $\boldsymbol{J} \times \boldsymbol{B}$ forces across this barrier effectively separate the ionosphere from the solar wind flow[11,35]. Our observations highlight the $\pm Y_{MSE}$ hemisphere asymmetry of the $J_{In}$. The connection between $J_{In}$ and magnetospheric currents appears more pronounced in the $-Y_{MSE}$ hemisphere, where $J_{In}$ predominantly flows electric northward. This suggests that the magnetic pressure gradient force at ionospheric altitude is likely stronger at the $-Y_{MSE}$ terminator and weaker at the $+Y_{MSE}$ terminator. Additionally, the tailward current in the terminator region may play a role in maintaining the nightside magnetospheric currents, although this warrants further investigation[67,68,69].

The $J_{Sq}$ current system is primarily concentrated on the dayside and terminator of Mars. This current system includes a tailward component in the terminator regions, generating a rotation of the magnetic field (Fig. 3e). The two horizontal current flows produce an inward/outward magnetic field in the Northern/Southern hemisphere, respectively, as observed in the MSO coordinate (Fig. 3d). The horizontally distributed magnetic field acts to deflect and shield the ionosphere from the solar wind plasma penetration, and preventing the escape of ionospheric particles. When observed at the Martian surface, apart from the daily variations of $B_\theta$, $B_\varphi$ components, $J_{Sq}$ also leads to daily variations in the $B_r$ component of the surface magnetic field.

### 3.4 Implications for other planets: Venus, Titan, and exoplanets

$J_{In}$ and $J_{Sq}$ current systems could be universal physical phenomena in planetary ionosphere. The $J_{In}$ current system is driven by the electric potential difference at bow shock and IMB, which occurs when solar wind encounters an obstacle in space, such as a planet's ionosphere. The $J_{Sq}$ current system arises from the differential movement of ions and electrons within a planetary ionosphere and is evident on planets with significant atmospheric tides. These current systems would arise when the stellar wind or magnetospheric plasma interacts with the ionospheres of unmagnetized planets, including but not limited to Venus, Titan, Pluto, and exoplanets[4,70,71,72].

Venus and Titan are ideal planets for studying ionospheric current systems due to their thick atmospheres and absence of intrinsic global magnetic fields. Previous missions, such as Venus Express and Cassini, have identified ionospheric magnetic fields on these planets, but the mechanisms driving these phenomena remain under debate[73,74]. Recent studies proposed that electric current controlled by the IMF direction might occur at ionospheric altitude on Venus[15,75]. The distribution of ionospheric currents on these planets deserves further examination to determine whether they are consistent with influences from the solar wind or neutral winds. Furthermore, our findings have implications for understanding the space environment on ancient Earth, particularly during periods of geomagnetic reversals or excursions when the geomagnetic field strength is very weak[76]. During these times, Earth's weakened geomagnetic field allowed direct solar wind-ionosphere interactions, forming an induced magnetosphere similar to that of Mars[77]. In these scenarios, the magnetic fields generated by current systems act as a shield against energetic cosmic and solar particles, preserving Earth's atmosphere and the life within. External magnetic fields become increasingly significant when the geomagnetic field is extremely weak, and evidence of rapidly changing magnetic field signatures may already have been recorded in paleomagnetic records[78,79].

## 4 Method

### 4.1 Data and Coordinates

This study is based on the magnetic field and plasma data obtained from Mars Atmosphere and Volatile EvolutioN (MAVEN) spacecraft, from November 1, 2014, to May 15, 2022. The orbit of MAVEN is eccentric, with an apoapsis of ~6200 km and a periapsis of ~120 km[24]. Following an aerobraking maneuver in 2019, the apoapsis altitude was adjusted to ~4400 km, resulting in a decreased orbital period. The period of the MAVEN orbit is ~4.5 hours, and its periapsis precesses in local time, resulting in a data set that covers all local times at ionosphere altitudes (120-500 km). The magnetic field data was measured by fluxgate magnetometers (MAG) onboard MAVEN[80], and the vector magnetic field data was resampled at a rate of 4 s. The velocity of the solar wind was measured by the Solar Wind Ion Analyzer (SWIA) instrument[81]. The electron density and electron temperature in the ionosphere were measured by the Langmuir Probe and Waves (LPW) instrument[82].

We analyzed the distribution of magnetic field and currents in both MSO and MSE coordinates. To rotate the data from MSO to MSE coordinates, we first evaluate upstream solar wind conditions. From a total of 13955 orbits, 8742 crossings were selected that fulfilled the criteria of having solar wind interval between neighboring crossings (between outbound and inbound) of the bow shock exceeding 1.5 hours. The bow shock crossing time was determined through a manual examination of observations, similar to previous procedures[48,83]. The direction of the solar wind flow and the IMF is determined by using 25-minutes averages of solar wind observations, starting 5 minutes after (before) the outbound (inbound) bow shock crossings, respectively. The resulting solar wind conditions for each orbit were then calculated as the average of these two observations, with the solar wind velocity given by $\boldsymbol{v}_{SW} = (\boldsymbol{v}_{SW1} + \boldsymbol{v}_{SW2})/2$ and the IMF by $\boldsymbol{B}_{IMF} = (\boldsymbol{B}_{IMF1} + \boldsymbol{B}_{IMF2})/2$. Furthermore, to ensure a stable IMF condition, we selected orbits which have the angle between $\boldsymbol{B}_{IMF1}$ and $\boldsymbol{B}_{IMF2}$ less than 45 degrees and the difference in the magnitude of $\boldsymbol{B}_{IMF1}$ and $\boldsymbol{B}_{IMF2}$, expressed as $\frac{|\boldsymbol{B}_{IMF1}|-|\boldsymbol{B}_{IMF2}|}{|\boldsymbol{B}_{IMF1}|+|\boldsymbol{B}_{IMF2}|}$, less than 0.2. Out of the 8742 orbits, 3103 orbits satisfied these criteria and 22% (3,103 out of 13,955) of the total data has been used in the statistics. Because upstream IMF conditions are constantly changing, the induced magnetosphere response to short-period IMF variations (<4.5-hour, period of the MAVEN orbit), such as IMF discontinuities[84], are smoothed out in the statistics.

Magnetic field distributions in the Martian ionosphere are essential for calculating ionospheric currents; however, it can be distorted by the localized crustal magnetic field (see Supplementary Fig. 16). We focused our analysis on ionospheric magnetic fields that are not influenced by the Martian crustal magnetic field. To achieve this, we only keep data measured in non-crustal field regions, defined as regions in which the intensity of the crustal magnetic fields at 120 km altitude $|B_{model,120\ km}|<10$ nT. In the Southern hemisphere, 20% of the data from non-crustal field regions remains. By testing different models to identify non-crustal field regions, including the L134[85] and G110[66] model, we find that our results regarding the current distribution are insensitive to the choice of crustal field model (Supplementary Fig. 17), but notable discrepancies in magnitude exist in the noon regions.

**4.2 Calculating the current and electric field**

We calculate the current density by taking the curl of the magnetic field. Considering Maxwell-Ampere's law $\nabla \times \boldsymbol{B} = \mu_0 \boldsymbol{J} + \varepsilon_0 \mu_0 \frac{\partial \boldsymbol{E}}{\partial t}$ and ignoring the displacement current term $\varepsilon_0 \mu_0 \frac{\partial \boldsymbol{E}}{\partial t}$, the current $\boldsymbol{J}$ can be calculated as $\boldsymbol{J} = \frac{1}{\mu_0} \nabla \times \boldsymbol{B}$. This method has been widely applied in studying current distributions in the planetary magnetosphere[11,86,87] and ionosphere[62].

In the magnetosphere, the spatial domain was set as $-3\ R_m < X_{MSE} < 3\ R_m$, $-3\ R_m < Y_{MSE} < 3\ R_m$, and $-3\ R_m < Z_{MSE} < 3\ R_m$ in the MSE Cartesian coordinate system, and was divided into bins with a width of $0.2\ R_m$, corresponding to 670 km. We only retained the bins for which the number of data points exceeded ten to decrease the statistical bias. In each bin, we calculated the average magnetic field as the mean value of all data points, while the current density at the center of each bin was calculated as

$$\boldsymbol{J} = (J_x, J_y, J_z) = \frac{1}{\mu_0}\left(\frac{\partial B_z}{\partial y} - \frac{\partial B_y}{\partial z}, \frac{\partial B_x}{\partial z} - \frac{\partial B_z}{\partial x}, \frac{\partial B_y}{\partial x} - \frac{\partial B_x}{\partial y}\right) \quad (1)$$

In the ionosphere, the calculation of the current density is more convenient in local spherical coordinates $(\mathbf{r}, \boldsymbol{\theta}, \boldsymbol{\varphi})$, defined within the frame of both the MSE and MSO coordinates, where $B_r$ points outward, $B_\theta$ points southward, and $B_\varphi$ points eastward. The current density is then calculated as:

$$\boldsymbol{J} = \begin{pmatrix} J_r, \\ J_\theta, \\ J_\varphi \end{pmatrix} = \frac{1}{\mu_0} \begin{pmatrix} \frac{1}{r\sin\theta}\frac{\partial}{\partial\theta}(B_\varphi \sin\theta) - \frac{1}{r\sin\theta}\frac{\partial B_\theta}{\partial\varphi}, \\ \frac{1}{r\sin\theta}\frac{\partial B_r}{\partial\varphi} - \frac{1}{r}\frac{\partial}{\partial r}(rB_\varphi), \\ \frac{1}{r}\frac{\partial}{\partial r}(rB_\theta) - \frac{1}{r}\frac{\partial B_r}{\partial\theta} \end{pmatrix} \quad (2)$$

where $r$ is the radial distance from the center of Mars, $\theta$ is the colatitude, and $\varphi$ is the longitude. The spatial domain is defined as a spherical shell extending from an altitude of 150 to 500 km and partitioned into bins of 18° latitude × 18° longitude × 50 km. For most regions, there are over 100 data points per bin in both MSO and MSE coordinates collected in the non-crustal field regions (Supplementary Fig. 1). The only significant data gap is located at the subsolar point in the MSE coordinates, due to MAVEN's highly eccentric orbit.

The derivatives of the magnetic field are calculated using the central difference, except at the boundaries of the spatial domain where the single-sided difference is used. Furthermore, given the uncertainties in ionospheric statistical results, we applied a 2-by-2 kernel low-pass filter to smooth the data. The physical error in the current calculation can be estimated by $|\nabla \cdot \mathbf{B}|/|\nabla \times \mathbf{B}|$[88] as shown in Supplementary Fig. 18. In general, the physical error of the ionospheric current is less than 0.1, indicating that the current calculation is physically correct.

The ionospheric electric field can be derived from ionospheric currents and conductivity. In the ionospheric dynamo region, the electric forces dominate over the gravity and the pressure gradient forces. Ohm's law defines the current density **J** as $\mathbf{J} = \sigma(\mathbf{E} + \mathbf{U} \times \mathbf{B})$. In the Martian ionosphere, the term $\mathbf{U} \times \mathbf{B}$ is considered negligible compared to **E** (Supplementary Fig. 13). The ionospheric electric field is related to the ionospheric currents as:

$$\mathbf{J} = \sigma_\parallel \mathbf{E}_\parallel + \sigma_P \mathbf{E}_\perp + \sigma_H \hat{b} \times \mathbf{E}_\perp \tag{3}$$

where the $\sigma_\parallel$, $\sigma_P$, and $\sigma_H$ represent the Parallel, Pedersen, and Hall conductivities, respectively. $\mathbf{E}_\parallel$ and $\mathbf{E}_\perp$ represent the electric field parallel and perpendicular to the local magnetic field. The $\hat{b}$ represent the unit vector along the local magnetic field direction. After ignoring the ion contribution to the parallel conductivity[89], ionospheric conductivities are determined as follows:

$$\begin{aligned}\sigma_\parallel &= \frac{n_e e^2}{m_e \nu_{en}} \\ \sigma_P &= \sum_i \frac{n_i e_i^2 \nu_{in}}{m_i(\nu_{in}^2 + \omega_i^2)} + \frac{n_e e^2 \nu_{en}}{m_e(\nu_{en}^2 + \omega_e^2)} \\ \sigma_H &= \frac{n_e e^2 \omega_e}{m_e(\nu_{en}^2 + \omega_e^2)} - \sum_i \frac{n_i e_i^2 \omega_i}{m_i(\nu_{in}^2 + \omega_i^2)}\end{aligned} \tag{4}$$

Here, $\nu_{in}$ ($\nu_{en}$) represents the collision frequency between ions (electrons) and each neutral species, over all neutral species, $\nu_i = \sum_n \nu_{in}$ ($\nu_e = \sum_n \nu_{en}$), respectively. $\omega_i$ ($\frac{|e|\cdot|\vec{B}|}{m_i}$) and $\omega_e$ ($\frac{|e|\cdot|\vec{B}|}{m_e}$) are the gyrofrequencies of ions and electrons, and $m_i$ and $m_e$ are the mass of ions and electron, respectively. The $n_e$ is the electron density, $e$ is the elementary charge, and $e_i$ is the quantity of charge of ions. It should be noted that electron-ion collision frequencies have been neglected in this method; however, they can be incorporated in more general cases[23].

In the Martian ionosphere, $v_{en}$ is primarily attributed to the collision frequency between electrons and neutral $CO_2$, and $v_{in}$ is primarily attributed to the collision frequency between $O_2^+$ and neutral $CO_2$ (table 4.4 and 4.6 in 90). They are calculated as

$$v_{en} = 3.68 * 10^{-8} n(CO_2)(1 + 4.1 \times 10^{-11}|4500 - T_e|^{2.93})$$
$$v_{in} = C_{in} n_n \tag{5}$$

The electron temperature, $T_e$, derived from the LPW instrument, is shown in Supplementary Figure 19d. The median value from the data is used in the conductivity calculations. The $CO_2$ density, denoted as $n(CO_2)$, is sourced from Figure 2.22 in 90. The $C_{in}$ is the collision coefficient for nonresonant ion–neutral collision, $C_{in} \times 10^{10} = 5.63$[90]. The magnetic field and electric density used in the conductivity calculations are derived from MAG and LPW instrument, as illustrated in Fig. 2 and Fig. 3 for the magnetic field, and in Supplementary Fig. 19a for electric density.

By solving equation (3), the perpendicular and parallel electric field are given by[16]

$$\boldsymbol{E}_\perp = \frac{\sigma_P}{\sigma_P^2 + \sigma_H^2}\boldsymbol{J}_\perp + \frac{\sigma_H}{\sigma_P^2 + \sigma_H^2}\boldsymbol{J}_\perp \times \hat{b}$$
$$\boldsymbol{E}_\parallel = \boldsymbol{J}_\parallel/\sigma_\parallel \tag{6}$$

where $\boldsymbol{J}_\perp$ and $\boldsymbol{J}_\parallel$ are the current perpendicular and parallel to the magnetic field, respectively. The ionosphere electric field is then calculated as $\boldsymbol{E} = \boldsymbol{E}_\perp + \boldsymbol{E}_\parallel$.

### 4.3 Multi-fluid Hall-MHD simulation

We employed a global 3D multi-fluid Hall-MHD (magnetohydrodynamic) simulation to corroborate our observations of the Martian currents[91,92]. The simulation solves the Navier–Stokes transport equations and the conservation equations for mass, momentum, and energy across four ion fluids: $H^+$, $O_2^+$, $O^+$, and $CO_2^+$, with details provided in Supplementary Text 1. Current density is derived from $\boldsymbol{J} = \frac{1}{\mu_0}\nabla \times \boldsymbol{B}$, consistent with the observation. The Martian ionosphere is self consistently created, incorporating solar flux ($F_{10.7}$) and optical depth calculated via the Chapman function. For simplicity, a 1D neutral density profile corresponding to solar maximum conditions serves as the initial input. The simulation adopts average solar wind parameters upstream of Mars as reported by 48; specifically, the solar wind density is 1.4 $cm^{-3}$, the solar wind velocity $(V_x, V_y, V_z) = (-367, 0, 0)$ km/s, and IMF $(B_x, B_y, B_z) = (0, 2.8, 0)$ nT in the MSE coordinate.

Importantly, we neglect Martian crustal magnetic fields in our simulation, to solely focus on the magnetic fields and currents in the induced magnetosphere.

### 4.4 Averaged wind field from MGCM model

We compare the neutral wind field derived from the ionospheric current with the Mars global circulation model (MGCM) model[93]. The MGCM model reaches up to the exobase, the upper boundary of the thermosphere, approximately at an altitude of 300 km. The source code is accessible upon request from the Mars Global Climate Database (MCD) website (https://www-mars.lmd.jussieu.fr/mars/access.html). For our analyses, we employed version 5.2 of the MCD. The horizontal wind field data is calculated for every 15° solar longitude from 0° to 360°, every hour from 0 h to 24 h, and at intervals of 6° in latitude. The input parameter 'hireskey' is set to 0, indicating data interpolation from the GCM in a 5.625° ×3.75° grid. The input parameter 'Dust' is set to 1, indicating the average solar EUV conditions. The zonal and meridional winds are determined by averaging over both the geographic longitudes and solar longitude.

### 4.5 Uncertainties estimate

We quantified the uncertainties associated with our statistical results. While the uncertainty of magnetospheric current distributions has previously been addressed[11], our evaluation concentrates on the ionospheric magnetic field and current uncertainties in the Martian ionosphere. The magnetic field uncertainties within individual bins are calculated as:

$$\sigma_{B_r} = \sqrt{\frac{\sum_{i=1}^{n}(B_{r,i}-\overline{B_r})^2}{n-1}}$$
$$\sigma_{B_\theta} = \sqrt{\frac{\sum_{i=1}^{n}(B_{\theta,i}-\overline{B_\theta})^2}{n-1}} \quad (7)$$
$$\sigma_{B_\varphi} = \sqrt{\frac{\sum_{i=1}^{n}(B_{\varphi,i}-\overline{B_\varphi})^2}{n-1}}$$

Here, $\sigma_{B_r}$, $\sigma_{B_\theta}$, and $\sigma_{B_\varphi}$ represent the standard error for the magnetic field $B_r$, $B_\theta$, and $B_\varphi$ components, respectively. $n$ denotes the data count within each bin. $\overline{B}$ denotes the mean value of the magnetic field. Current uncertainties were derived through error propagation in equation 2[94], leading to:

$$\sigma_{J_r} = \frac{1}{\mu_0} \sqrt{\left(\frac{1}{r} * \sigma_{\frac{\Delta B_\varphi}{\Delta \theta}}\right)^2 + \left(\frac{\cos\theta}{r\sin\theta} * \sigma_{B_\varphi}\right)^2 + \left(\frac{1}{r\sin\theta} * \sigma_{\frac{\Delta B_\theta}{\Delta \varphi}}\right)^2}$$

$$\sigma_{J_\theta} = \frac{1}{\mu_0} \sqrt{\left(\frac{1}{r\sin\theta} * \sigma_{\frac{\Delta B_r}{\Delta \varphi}}\right)^2 + \left(\sigma_{\frac{\Delta B_\varphi}{\Delta r}}\right)^2 + \left(\frac{1}{r}\sigma_{B_\varphi}\right)^2} \qquad (8)$$

$$\sigma_{J_\varphi} = \frac{1}{\mu_0} \sqrt{\left(\sigma_{\frac{\Delta B_\theta}{\Delta r}}\right)^2 + \left(\frac{1}{r}\sigma_{B_\theta}\right)^2 + \left(\frac{1}{r} * \sigma_{\frac{\Delta B_r}{\Delta \theta}}\right)^2}$$

To calculate the current uncertainty, we have to first estimate the uncertainty of the magnetic field gradient. For example, the uncertainty of the $B_\theta$ gradient in the radial direction based on the first-order Taylor expansion, for the central difference method, is given by:

$$\sigma_{\frac{\Delta B_\theta}{\Delta r}} = \frac{\sqrt{\sigma_{B_\theta}^2(r_{i-1},\theta_j,\varphi_k) + 2\sigma_{B_\theta}^2(r_i,\theta_j,\varphi_k) + \sigma_{B_\theta}^2(r_{i+1},\theta_j,\varphi_k)}}{\sqrt{2}\Delta r} \qquad (9)$$

where $\Delta r$ denotes the bin size in the radial direction. Similar expressions can be constructed for the uncertainties related to the magnetic field gradient in other directions. Supplementary Fig. 20-23 show uncertainties of the magnetic field and currents at ionospheric altitudes. It is evident that the uncertainties, both for the magnetic fields and currents, are larger on the dayside, predominantly in horizontal components. Specifically, the $\sigma_{B_\theta}$ can rise up to 20 nT at the subsolar point at an altitude of 150 km. Although the magnitude of $J_r$ in Fig. 2j is relatively low and comparable to the uncertainty level, its extensive coverage across a wide range of longitudes in the equatorial region enhances its reliability. The substantial uncertainties in the magnetic field and current measurements indicate that ionospheric currents are influenced by both, the solar wind and the neutral wind, implying that no single coordinate frame can capture their dynamics. For statistical significance, 94.6% (76.1%), 95.3% (71.8%), and 95.4% (70.8%) of the magnetic field $B_r$, $B_\theta$, and $B_\varphi$ values at 150 km altitude are determined at $2\sigma$ ($1\sigma$) confidence, respectively.

Furthermore, the ionospheric electric field can only be estimated qualitatively, given the significant fluctuations in electron temperature and electron density of the ionosphere. The differences between the upper and lower quartiles (25% and 75%) of these values are quite pronounced (Supplementary Fig. 19), with more than one order of magnitude between them. These observations align with previous statistical findings that the electron density and electron temperature are highly perturbated[95,96].

**Data Availability**

All MAVEN data used in this paper are available from NASA's Planetary Data System (https://pds.nasa.gov/). MAG data can be found at https://pds-ppi.igpp.ucla.edu/mission/MAVEN/MAVEN/MAG. SWIA data can be found at https://pds-ppi.igpp.ucla.edu/mission/MAVEN/MAVEN/SWIA. LPW data is available at https://pds-ppi.igpp.ucla.edu/mission/MAVEN/MAVEN/LPW. The source data files for the figures have been deposited in a Zenodo repository https://zenodo.org/records/12789985.

**Code Availability**

The code to reproduce the figures are available at https://github.com/gaojiawei321/Mars_ionosphere_current.

**Author Contributions**

J. Gao and Z. Rong conceived this study. J. Gao carried out the MAVEN data analysis and lead the manuscript preparation. S. Li and H. Lu performed the Hall-MHD simulation. A. Mittelholz carried out the statistical work using the MGCM model. Z. Shi plotted Figure 1a. J. Gao, A. Mittelholz, Z. Rong, M. Persson, Z. Shi, C. Zhang, X. Wang, Y. Pan revised the manuscript. All authors contributed to the discussion and commented on the manuscript.

**Acknowledgments**

This work is supported by the National Natural Science Foundation of China (Grant No. 42304186), China Postdoctoral Science Foundation (2023M743466), the Strategic Priority Research Program of Chinese Academy of Sciences (Grant No. XDA17010201), the Key Research Program of Chinese Academy of Sciences (Grant No. ZDBS-SSW-TLC00103), the Key Research Program of the Institute of Geology & Geophysics, CAS (Grant No. IGGCAS- 201904, IGGCAS-202102), and the SNSF Ambizione fellowship. This research was supported by the International Space Science Institute (ISSI) in Bern and Beijing, through ISSI/ISSI-BJ International Team project "Understanding the Mars Space Environment through Multi-Spacecraft Measurements" (ISSI Team project #23–582; ISSI-BJ Team project #58). We would like to thank the entire MAVEN team and instrument leads for access to data and support. We thank Xinzhou Li, Yosuke Yamazaki, Zhaopeng Wu, Rentong Lin and Qi Zhang for their helpful discussion.

**Competing interests**

The authors declare no competing interests.

**References**

1. Luhmann, J. G., Ledvina, S. A., & Russell, C. T. (2004). Induced magnetospheres. Advances in Space Research, 33(11), 1905–1912. https://doi.org/10.1016/j.asr.2003.03.031
2. Nagy, A. F., Winterhalter, D., Sauer, K., Cravens, T. E., Brecht, S., Mazelle, C., et al. (2004). The plasma environment of Mars. Space Science Reviews, 111(1/2), 33–114. https://doi.org/10.1023/B:SPAC.0000032718.47512.92

3. Brain, D., Barabash, S., Bougher, S., Duru, F., Jakosky, B., & Modolo, R. (2017). Solar Wind Interaction and Atmospheric Escape. In R. Haberle, R. Clancy, F. Forget, M. Smith, & R. Zurek (Eds.), The Atmosphere and Climate of Mars (Cambridge Planetary Science, pp. 464-496). Cambridge: Cambridge University Press. doi:10.1017/9781139060172.015
4. Bertucci, C., Duru, F., Edberg, N., Fraenz, M., Martinecz, C., Szego, K., & Vaisberg, O. (2011). The induced magnetospheres of Mars, Venus, and Titan. Space Science Reviews, 162(1–4), 113–171. https://doi.org/10.1007/s11214-011-9845-1
5. Futaana, Y., Stenberg Wieser, G., Barabash, S., & Luhmann, J. G. (2017). Solar wind interaction and impact on the Venus atmosphere. Space Science Reviews, 212(3–4), 1453–1509. https://doi.org/10.1007/s11214-017-0362-8
6. Dong, C., Jin, M., & Lingam, M. (2020). Atmospheric escape from TOI-700 d: Venus versus Earth analogs. The Astrophysical Journal Letters, 896(2), L24
7. Barabash, S., Fedorov, A., Lundin, R., & Sauvaud, J. A. (2007). Martian atmospheric erosion rates. Science, 315(5811), 501-503
8. Lillis, R. J., Brain, D. A., Bougher, S. W., Leblanc, F., Luhmann, J. G., Jakosky, B. M., ... & Lin, R. P. (2015). Characterizing atmospheric escape from Mars today and through time, with MAVEN. Space Science Reviews, 195, 357-422.
9. Jakosky, B. M., Brain, D., Chaffin, M., Curry, S., Deighan, J., Grebowsky, J., ... & Zurek, R. (2018). Loss of the Martian atmosphere to space: Present-day loss rates determined from MAVEN observations and integrated loss through time. Icarus, 315, 146-157.
10. Baumjohann, W., M. Blanc, A. Fedorov, and K.-H. Glassmeier (2010), Current systems in planetary magnetospheres and ionospheres, Space Sci. Rev., 152, 99–134, doi:10.1007/s11214-010-9629-z
11. Ramstad, R., Brain, D. A., Dong, Y., Espley, J., Halekas, J., & Jakosky, B. (2020). The global current systems of the Martian induced magnetosphere. Nature Astronomy, 4(10), 979-985. https://doi.org/10.1038/s41550-020-1099-y
12. Chapman, S., & Ferraro, V. C. A. (1930). A new theory of magnetic storms. Nature, 126, 129–130. https://doi.org/10.1038/126129a0
13. Axford, W. I., Petschek, H. E., & Siscoe, G. L. (1965). Tail of the magnetosphere. Journal of Geophysical Research, 70(5), 1231-1236.
14. Wang, X. D., Fatemi, S., Holmström, M., Nilsson, H., Futaana, Y., & Barabash, S. (2024). Martian global current systems and related solar wind energy transfer: hybrid simulation under nominal conditions. Monthly Notices of the Royal Astronomical Society, 527(4), 12232-12242.
15. Dubinin, E., M. Fraenz, T. L. Zhang, J. Woch, and Y. Wei (2014), Magnetic fields in the Venus ionosphere: Dependence on the IMF direction—Venus express observations, J. Geophys. Res. Space Physics, 119, doi:10.1002/2014JA020195
16. Fillingim, M. O. (2018). Ionospheric currents at Mars and their electrodynamic effects. In Electric Currents in Geospace and Beyond, (Vol. 235, pp. 445–458). Hoboken, NJ, USA: John Wiley & Sons, Inc. https://doi.org/10.1002/9781119324522
17. Brain, D., F. Bagenal, M. H. Acuña, and J. Connerney (2003), Martian magnetic morphology: Contributions from the solar wind and crust, J. Geophys. Res., 108(A12), 1424, doi:10.1029/2002JA009482
18. Fillingim, M., L. Peticolas, R. Lillis, D. Brain, J. Halekas, D. Lummerzheim, and S. Bougher (2010), Localized ionization patches in the nighttime ionosphere of Mars and their electrodynamic consequences, Icarus, 206(1), 112–119.

19. Hamil, O., Cravens, T. E., Renzaglia, A., & Andersson, L. (2022). Small Scale Magnetic Structure in the Induced Martian Ionosphere and Lower Magnetic Pile-Up Region. Journal of Geophysical Research: Space Physics, 127(4), e2021JA030139.
20. Opgenoorth, H. J., Dhillon, R. S., Rosenqvist, L., Lester, M., Edberg, N. J. T., Milan, S. E., ... & Brain, D. (2010). Day-side ionospheric conductivities at Mars. Planetary and Space Science, 58(10), 1139-1151.
21. Riousset, J. A., C. S. Paty, R. J. Lillis, M. O. Fillingim, S. L. England, P. G. Withers, and J. P. M. Hale (2013), Three-dimensional multifluid modeling of atmospheric electrodynamics in Mars' dynamo region, J. Geophys. Res., 118, 3647–3659; doi:10.1002/jgra.50328.
22. Mittelholz, A., C. L. Johnson, and R. J. Lillis (2017), Global-scale external magnetic fields at Mars measured at satellite altitude, J. Geophys. Res. Planets, 122, 1243–1257, doi:10.1002/2017JE005308.
23. Lillis, R. J., Fillingim, M. O., Ma, Y., Gonzalez-Galindo, F., Forget, F., Johnson, C. L., ... & Fowler, C. M. (2019). Modeling wind-driven ionospheric dynamo currents at Mars: Expectations for InSight magnetic field measurements. Geophysical Research Letters, 46(10), 5083-5091.
24. Jakosky, B. M., Lin, R. P., Grebowsky, J. M., Luhmann, J. G., Mitchell, D. F., Beutelschies, G., ... & Zurek, R. (2015). The Mars atmosphere and volatile evolution (MAVEN) mission. Space Science Reviews, 195, 3-48.
25. Johnson, C. L., Mittelholz, A., Langlais, B., Russell, C. T., Ansan, V., Banfield, D., et al. (2020). Crustal and time-varying magnetic fields at the InSight landing site on Mars. Nature Geoscience, 13(3), 199–204. https://doi.org/10.1038/s41561-020-0537-x
26. Du, A., Ge, Y., Wang, H., Li, H., Zhang, Y., Luo, H., ... & Zhang, K. (2023). Ground magnetic survey on Mars from the Zhurong rover. Nature Astronomy, 1-11.
27. Mittelholz, A., Johnson, C. L., Thorne, S. N., Joy, S., Barrett, E., Fillingim, M. O., et al. (2020). The origin of observed magnetic variability for a sol on Mars from InSight. Journal of Geophysical Research: Planets, 125(9), 1–14. https://doi.org/10.1029/2020JE006505
28. Boscoboinik, G., Bertucci, C., Gomez, D., Dong, C., Regoli, L., Mazelle, C., ... & Andersson, L. (2023). Forces, electric fields and currents at the subsolar martian MPB: MAVEN observations and multifluid MHD simulation. Icarus, 401, 115598.
29. Acuña, M. H., et al. (1998), Magnetic field and plasma observations at Mars: Initial results of the Mars global surveyor mission, Science, 279, 1676–1680, doi:10.1126/Science.279.5357.1676.
30. Connerney, J. E. P., Acuna, M. H., Wasilewski, P. J., Ness, N. F., Rème, H., Mazelle, C., et al. (1999). Magnetic lineations in the ancient crust of Mars. Science, 284(5415), 794–798. https://doi.org/10.1126/science.284.5415.794
31. Weber, T., Brain, D., Xu, S., Mitchell, D., Espley, J., Halekas, J., et al. (2020). The influence of interplanetary magnetic field direction on Martian crustal magnetic field topology. Geophysical Research Letters, 47(19), e2020GL087757. https://doi.org/10.1029/2020GL087757
32. Withers, P., Mendillo, M., Rishbeth, H., Hinson, D. P., & Arkani-Hamed, J. (2005). Ionospheric characteristics above Martian crustal magnetic anomalies. Geophysical Research Letters, 32, L16204. https://doi.org/10.1029/2005GL023483
33. Fillingim, M. O., R. J. Lillis, S. L. England, L. M. Peticolas, D. A. Brain, J. S. Halekas, C. Paty, D. Lummerzheim, and S. W. Bougher (2012), On wind-driven electrojets at magnetic cusps in the nightside ionosphere of Mars, Earth Planets Space, 64, 93–103; doi:10.5047/eps.2011.04.010.
34. Riousset, J. A., C. S. Paty, R. J. Lillis, M. O. Fillingim, S. L. England, P. G. Withers, and J. P. M. Hale (2014), Electrodynamics of the Martian dynamo region near magnetic cusps and loops, Geophys. Res. Lett., 41, 1119‐1125; doi:10.1002/2013GL059130.
35. Halekas, J. S., Brain, D. A., Luhmann, J. G., DiBraccio, G. A., Ruhunusiri, S., Harada, Y., … Jakosky, B. M. (2017a). Flows, fields, and forces in the Mars-solar wind interaction. Journal of Geophysical Research: Space Physics, 122, 11,320–11,341. https://doi.org/10.1002/2017JA024772
36. Halekas, J. S., Ruhunusiri, S., Harada, Y., Collinson, G., Mitchell, D. L., Mazelle, C., et al. (2017b). Structure, dynamics, and seasonal variability of the mars-solar wind interaction: MAVEN Solar Wind Ion Analyzer in-flight performance and science results. Journal of Geophysical Research: Space Physics, 122(1), 547–578. https://doi.org/10.1002/2016JA023167
37. Zhang, C., Rong, Z., Klinger, L., Nilsson, H., Shi, Z., He, F., et al. (2022). Three-dimensional configuration of induced magnetic fields around Mars. Journal of Geophysical Research: Planets, 127, e2022JE007334. https://doi.org/10.1029/2022JE007334
38. McComas, D. J., Spence, H. E., Russell, C. T., & Saunders, M. A. (1986). The average magnetic field draping and consistent plasma properties in the Venus magnetotail. Journal of Geophysical Research, 91(A7), 7939–7953. https://doi.org/10.1029/JA091iA07p07939
39. Zhang, T., Baumjohann, W., Du, J., Nakamura, R., Jarvinen, R., Kallio, E., et al. (2010). Hemispheric asymmetry of the magnetic field wrapping pattern in the Venusian magnetotail. Geophysical Research Letters, 37(14), L14202. https://doi.org/10.1029/2010GL044020
40. Du, J., Wang, C., Zhang, T. L., & Kallio, E. (2013). Asymmetries of the magnetic field line draping shape around Venus. Journal of Geophysical Research: Space Physics, 118(11), 6915-6920.
41. Chai, L., Wan, W., Wei, Y., Zhang, T., Exner, W., Fraenz, M., et al. (2019). The induced global looping magnetic field on Mars. The Astrophysical Journal Letters, 871(2), L27. https://doi.org/10.3847/2041-8213/aaff6e
42. Dubinin, E., Modolo, R., Fraenz, M., Päetzold, M., Woch, J., Chai, L., et al. (2019). The induced magnetosphere of Mars: Asymmetrical topology of the magnetic field lines. Geophysical Research Letters, 46(22), 12722–12730. https://doi.org/10.1029/2019gl084387
43. Dubinin, E., Fraenz, M., Pätzold, M., Tellmann, S., Modolo, R., DiBraccio, G., et al. (2023). Magnetic fields and plasma motions in a hybrid Martian magnetosphere. Journal of Geophysical Research: Space Physics, 128, e2022JA030575. https://doi.org/10.1029/2022JA030575
44. Fang, X., Ma, Y., Luhmann, J., Dong, Y., Halekas, J., & Curry, S. (2023). Mars global distribution of the external magnetic field and its variability: MAVEN observation and MHD prediction. Journal of Geophysical Research: Space Physics, 128, e2023JA031588. https://doi.org/10.1029/2023JA031588
45. Ferguson, B. B., Cain, J. C., Crider, D. H., Brain, D. A., & Harnett, E. M. (2005). External fields on the nightside of Mars at Mars Global Surveyor mapping altitudes. Geophysical Research Letters, 32, L16105. https://doi.org/10.1029/2004GL021964
46. Fowler, C. M., Lee, C. O., Xu, S., Mitchell, D. L., Lillis, R., Weber, T., et al. (2019). The penetration of draped magnetic field into the Martian upper ionosphere and correlations with upstream solar wind dynamic pressure. Journal of Geophysical Research: Space Physics,124,3021–3035. https://doi.org/10.1029/2019JA026550
47. Huang, J. P., Hao, Y. Q., Lu, H. Y., & Cui, J. (2023). Variability of Draped Interplanetary Magnetic Field in the Subsolar Martian Ionosphere. The Astrophysical Journal, 955(1), 48.
48. Liu, D., Rong, Z., Gao, J., He, J., Klinger, L., Dunlop, M. W., et al. (2021). Statistical properties of solar wind upstream of Mars: MAVEN observations. The Astrophysical Journal, 911(2), 113. https://doi.org/10.3847/1538-4357/abed50
49. Xu, S., Mitchell, D. L., Mcfadden, J. P., Collinson, G., Harada, Y., Lillis, R., et al. (2018). Field-aligned potentials at Mars from MAVEN observations. Geophysical Research Letters, 45, 10,119–10,127. https://doi.org/10.1029/2018GL080136
50. Luhmann, J. G., Dong, C., Ma, Y., Curry, S. M., Mitchell, D., Espley, J., et al. (2015). Implications of MAVEN Mars near-wake measurements and models. Geophysical Research Letters, 42(21), 9087–9094. https://doi.org/10.1002/2015gl066122
51. Wang, X. D., Fatemi, S., Nilsson, H., Futaana, Y., Holmström, M., & Barabash, S. (2023). Solar wind interaction with Mars: electric field morphology and source terms. Monthly Notices of the Royal Astronomical Society, 521(3), 3597-3607.
52. Zhang, Q., Holmström, M., & Wang, X. (2023). Effects of ion composition on escape and morphology at Mars. Annales Geophysicae, 41(2), 375–388. https://doi.org/10.5194/angeo-41-375-2023
53. Mendillo, M., et al. (2017), MAVEN and the total electron content of the Martian ionosphere, J. Geophys. Res. Space Physics, 122, 3526-3537, doi:10.1002/2016JA023474
54. Andrews, D. J., Stergiopoulou, K., Andersson, L., Eriksson, A. I. E., Ergun, R. E., & Pilinski, M. (2023). Electron densities and temperatures in the Martian ionosphere: MAVEN LPW observations of control by crustal fields. Journal of Geophysical Research: Space Physics, 128, e2022JA031027. https://doi.org/10.1029/2022JA031027
55. Benna, M., Bougher, S. W., Lee, Y., Roeten, K. J., Yiğit, E., Mahaffy, P. R., & Jakosky, B. M. (2019). Global circulation of Mars' upper atmosphere. Science, 366(6471), 1363-1366.
56. Vasyliūnas, V. M. (2012). The physical basis of ionospheric electrodynamics. In Annales Geophysicae (Vol. 30, No. 2, pp. 357-369). Göttingen, Germany: Copernicus Publications.
57. Withers, P. (2008), Theoretical models of ionospheric electrodynamics and plasma transport, J. Geophys. Res., 113, A07301; doi:10.1029/2007JA012918.
58. Yamazaki, Y., & Maute, A. (2017). Sq and EEJ-A Review on the daily variation of the geomagnetic field caused by ionospheric dynamo currents. Space Science Reviews, 206, 299–405. https://doi.org/10.1007/s11214-016-0282-z
59. Takeda, M., & Araki, T. (1985). Electric conductivity of the ionosphere and nocturnal currents. Journal of atmospheric and terrestrial physics, 47(6), 601-609.
60. González-Galindo, F., Eusebio, D., Němec, F., Peter, K., Kopf, A., Tellmann, S., & Paetzold, M. (2021). Seasonal and geographical variability of the Martian ionosphere from Mars express observations. Journal of Geophysical Research: Planets, 126(2), e2020JE006661.
61. Mittelholz, A., Johnson, C. L., Fillingim, M., Grimm, R. E., Joy, S., Thorne, S. N., & Banerdt, W. B. (2023). Mars' external magnetic field as seen from the surface with InSight. Journal of Geophysical Research: Planets, 128, e2022JE007616. https://doi.org/10.1029/2022JE007616

62. Tozzi, R., M. Pezzopane, P. De Michelis, and M. Piersanti (2015), Applying a curl-B technique to Swarm vector data to estimate nighttime F region current intensities, Geophys. Res. Lett., 42, 6162–6169, doi:10.1002/2015GL064841.
63. Coxon, J. C., Milan, S. E., & Anderson, B. J. (2018). A review of Birkeland current research using AMPERE. Electric currents in geospace and beyond, 257-278.
64. Yamazaki, Y., Richmond, A.D. A theory of ionospheric response to upward-propagating tides: electrodynamic effects and tidal mixing effects. J. Geophys. Res. Space Phys. 118, 5891–5905 (2013). doi:10.1002/jgra.50487
65. Kleinböhl, A., John Wilson, R., Kass, D., Schofield, J. T., & McCleese, D. J. (2013). The semidiurnal tide in the middle atmosphere of Mars. Geophysical Research Letters, 40(10), 1952-1959.
66. Gao, J. W., Rong, Z. J., Lucy, K., Li, X. Z., Liu, D., & Wei, Y. (2021). A spherical harmonic Martian crustal magnetic field model combining data sets of MAVEN and MGS. Earth and Space Science, 8(10), e2021EA001860. https://doi.org/10.1029/2021EA001860
67. Xu, S., Mitchell, D., Liemohn, M., Dong, C., Bougher, S., Fillingim, M., Lillis, R., McFadden, J., Mazelle, C., Connerney, J., & Jakosky, B. (2016). Deep nightside photoelectron observations by MAVEN SWEA: Implications for Martian Northern hemispheric magnetic topology and nightside ionosphere source. Geophysical Research Letters, 43, 8876–8884. https://doi.org/10.1002/2016GL070527
68. Adams, D., Xu, S., Mitchell, D. L., Lillis, R. J., Fillingim, M., Andersson, L., et al. (2018). Using magnetic topology to probe the sources of Mars' nightside ionosphere. Geophysical Research Letters, 45, 12,190–12,197. https://doi.org/10.1029/2018GL080629
69. Cui, J., Cao, Y. T., Wu, X. S., Xu, S. S., Yelle, R. V., Stone, S., Vigren, E., Edberg, N. J. T., Shen, C. L., & He, F. (2019). Evaluating local ionization balance in the nightside Martian upper atmosphere during MAVEN Deep Dip campaigns. The Astrophysical Journal Letters, 876(1), L12. https://doi.org/10.3847/2041-8213/ab1b34
70. Chai, L., Wei, Y., Wan, W., Zhang, T., Rong, Z., Fraenz, M., et al. (2016). An induced global magnetic field looping around the magnetotail of Venus. Journal of Geophysical Research: Space Physics, 121(1), 688–698. https://doi.org/10.1002/2015ja021904
71. Dong, C., Lingam, M., Ma, Y., & Cohen, O. (2017). Is Proxima Centauri b habitable? A study of atmospheric loss. The Astrophysical Journal Letters, 837(2), L26.
72. Dong, C., Jin, M., Lingam, M., Airapetian, V. S., Ma, Y., & van der Holst, B. (2018). Atmospheric escape from the TRAPPIST-1 planets and implications for habitability. Proceedings of the National Academy of Sciences, 115(2), 260-265.
73. Bertucci, C., et al. (2008), The magnetic memory of Titan's ionized atmosphere, Science, 321(5895), 1475–1478
74. Zhang, T. L., Baumjohann, W., Russell, C. T., Villarreal, M. N., Luhmann, J. G., & Teh, W. L. (2015). A statistical study of the low‐altitude ionospheric magnetic fields over the north pole of Venus. Journal of Geophysical Research: Space Physics, 120(8), 6218-6229.
75. He, M., Vogt, J., Dubinin, E., Zhang, T., & Rong, Z. (2021). Spatially highly resolved solar-wind-induced magnetic field on Venus. The Astrophysical Journal, 923(1), 73.
76. Glassmeier, K. H., Vogt, J., 2010. Magnetic polarity transitions and biospheric effects. Space Sci. Rev. 155, 387–410.
77. Gong, F., Yu, Y., Cao, J., Wei, Y., Gao, J., Li, H., et al. (2022). Simulating the solar wind-magnetosphere interaction during the Matuyama-Brunhes paleomagnetic reversal. Geophysical Research Letters, 49(3), e97340. https://doi.org/10.1029/2021GL097340

78. Coe, R. S., Prevot, M., and Camps, P. (1995). New evidence for extraordinarily rapid change of the geomagnetic field during a reversal. Nature, 374:687–692.
79. Jackson, A. (1995). Storm in a lava flow? Nature, 377:685–686.
80. Connerney, J. E. P., J. Espley, P. Lawton, S. Murphy, J. Odom, R. Oliversen, and D. Sheppard (2015b), The MAVEN magnetic field investigation, Space Sci. Rev., doi:10.1007/s11214-015-0169-4.
81. Halekas, J. S., E. R. Taylor, G. Dalton, G. Johnson, D. W. Curtis, J. P. McFadden, D. L. Mitchell, R. P. Lin, and B. M. Jakosky (2013), The solar wind ion analyzer for MAVEN, Space Sci. Rev., doi:10.1007/s11214-013-0029-z.
82. Andersson, L., Ergun, R. E., Delory, G., Eriksson, A. I., Westfall, J., Reed, H., et al. (2015). The Langmuir probe and eaves experiment for MAVEN. Space Science Reviews, 195(1-4), 173–198. https://doi.org/10.1007/s11214-015-0194-3
83. Gruesbeck, J. R., Espley, J. R., Connerney, J. E. P., DiBraccio, G. A., Soobiah, Y. I., Brain, D., et al. (2018). The three-dimensional bow shock of Mars as observed by MAVEN. Journal of Geophysical Research: Space Physics, 123(6), 4542–4555. https://doi.org/10.1029/2018JA025366
84. Romanelli, N., DiBraccio, G., Modolo, R., Leblanc, F., Espley, J., Gruesbeck, J., et al. (2019). Recovery timescales of the dayside Martian magnetosphere to IMF variability. Geophysical Research Letters, 46, 10977–10986. https://doi.org/10.1029/2019GL084151
85. Langlais, B., Thébault, E., Houliez, A., Purucker, M. E., & Lillis, R. J. (2019). A new model of the crustal magnetic field of Mars using MGS and MAVEN. Journal of Geophysical Research: Planets, 124, 1542–1569. https://doi.org/10.1029/2018JE005854
86. Le, G., Russell, C. T., & Takahashi, K. (2004). Morphology of the ring current derived from magnetic field observations. Annales Geophysicae, 22(4), 1267–1295. https://doi.org/10.5194/angeo-22-1267-2004
87. Shi, Z., Rong, Z. J., Fatemi, S., Slavin, J. A., Klinger, L., Dong, C., et al. (2022). An eastward current encircling Mercury. Geophysical Research Letters, 49, e2022GL098415. https://doi.org/10.1029/2022GL098415
88. Robert, P., Dunlop, M. W., Roux, A., & Chanteur, G. E. R. A. R. D. (1998). Accuracy of current density determination. Analysis methods for multi-spacecraft data, 398, 395-418.
89. Takeda, M., & Araki, T. (1985). Electric conductivity of the ionosphere and nocturnal currents. Journal of atmospheric and terrestrial physics, 47(6), 601-609.
90. Schunk, R. W., & Nagy, A. (2009). Ionospheres: Physics, plasma physics, and chemistry, (2nd ed.). New York: Cambridge Univ. Press. https://doi.org/10.1017/CBO9780511635342
91. Li, S., Lu, H., Cui, J., Yu, Y., Mazelle, C., Li, Y., & Cao, J. (2020). Effects of a dipole‐like crustal field on solar wind interaction with Mars. Earth and Planetary Physics, 4(1), 23-31.
92. Li, S., Lu, H., Cao, J., Cui, J., Ge, Y., Zhang, X., ... & Wang, J. (2023). Global Electric Fields at Mars Inferred from Multifluid Hall-MHD Simulations. The Astrophysical Journal, 949(2), 88.
93. Millour, E., F. Forget, A. Spiga, L. Montabone, S. Lebonnois, S. R. Lewis, and P. L. Rea (2015), The Mars climate database (v4.3). EPSC Abstracts vol. 10, EPSC2015-438, paper presented at European Planetary Science Congress 2015, Nantes, France, 27 Sep.–2 Oct.
94. Bevington, P. R., & Robinson, D. K. (2003). Data reduction and error analysis. McGraw-Hill, New York.
95. Lillis, R. J., Fillingim, M. O., & Brain, D. A. (2011). Three-dimensional structure of the Martian nightside ionosphere: Predicted rates of impact ionization from Mars Global

Surveyor magnetometer and electron reflectometer measurements of precipitating electrons. Journal of Geophysical Research: Space Physics, 116(A12).
96. Vogt, M. F., et al. (2016), MAVEN observations of dayside peak electron densities in the ionosphere of Mars, J. Geophys. Res. Space Physics, 121, 891–906, doi:10.1002/2016JA023473
97. Němec, F., Linzmayer, V., Němeček, Z., & Šafránková, J. (2020). Martian bow shock and magnetic pileup boundary models based on an automated region identification. Journal of Geophysical Reseaarch: Space Physics, 125, e2020JA028509. https://doi.org/10.1029/2020JA028509